Stresa, Italy, 25-27 April 2007# TOWARDS A METHODOLOGY FOR ANALYSIS OF INTERCONNECT STRUCTURES FOR 3D-INTEGRATION OF MICRO SYSTEMS

*Peter Schneider, Sven Reitz, Andreas Wilde, Günter Elst, Peter Schwarz*

*Fraunhofer Institute for Integrated Circuits,*
*Branch Lab Design Automation, Dresden, Germany*
Email: peter.schneider@eas.iis.fraunhofer.de, Tel: +49 (351) 4640 710## ABSTRACT

Functional aspects as well as the influence of integration technology on the system behavior have to be considered in the 3D integration design process of micro systems. Therefore, information from different physical domains has to be provided to designers. Due to the variety of structures and effects of different physical domains, efficient modeling approaches and simulation algorithms have to be combined. The paper describes a modular approach which covers detailed analysis with PDE solvers and model generation for system level simulation.## 1. INTRODUCTION

New technologies for three-dimensional integration of sensors, electronics for signal processing and communication circuits enable a wide range of new applications of micro systems [1]. In Fig. 1 a typical structure including different interconnect technologies is shown.

Due to the dense integration a variety of physical effects have to be considered within the design process.

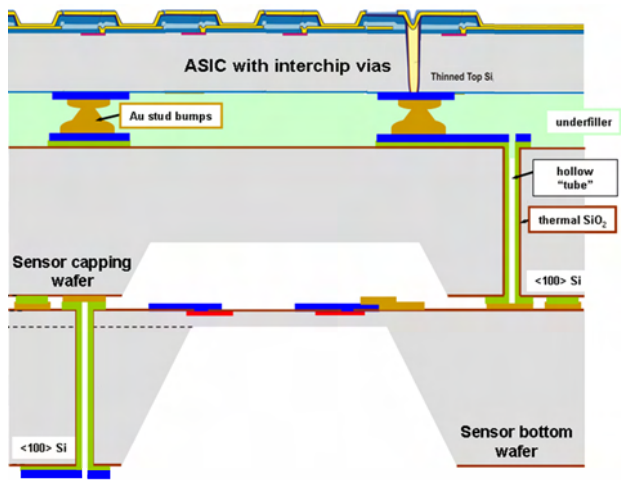

*Figure 1: Stacked micro system (contribution by SINTEF, Oslo, Norway)*

Basically, parasitics of semiconductor devices, the electrical behavior of interconnects, thermal interactions within the stacked structure as well as thermo-mechanical effects have to be investigated to meet requirements concerning high system performance and reliability [2], [3]. Due to the variety of physical effects and relevant design tools, an appropriate analysis methodology is required.

## 2. METHODOLOGY FOR MULTI-LEVEL AND MULTI-PHYSICS ANALYSIS OF INTERCONNECT STRUCTURES

Modeling of the mentioned effects leads to multi-physics models with a combination of electrical, electromagnetic, thermal, and even mechanical parts. Depending on the design step different models are required and different types of simulations have to be carried out (see Fig. 2).

For instance, thermo-mechanical analyses are used to investigate the reliability of single vias and small assemblies of such vias, respectively. Variations of geometry and material parameters can be analyzed [4], [5]. Performing electrostatic and electromagnetic calculations of single vias and local interconnect structures in the RF domain using PDE solvers, parasitic circuit elements like resistors, capacitors, and inductors can be extracted and later used to derive behavioral models for system level simulation. PDE solvers are also applied to investigate the thermal behavior of local interconnect structures and the stack assembly and to derive behavioral models. Electromagnetic and thermal behavioral models can be combined with the electrical model of the entire system to enable the complete system simulation considering the effects mentioned above.

A methodology, which complies with these prerequisites, comprises:

*1. Modular modeling approach*
This approach comprises a tool-independent structural representation, a semi-automatically generation of models for PDE solvers, detailed simulations with PDE solvers as well as the construction of behavioral models for system level simulation with adjustable accuracy [6]. The described flow and the assembly of complex models from basic models are supported by unified interfaces.

©EDA Publishing/DTIP 2007                                                  ISBN: 978-2-35500-000-3

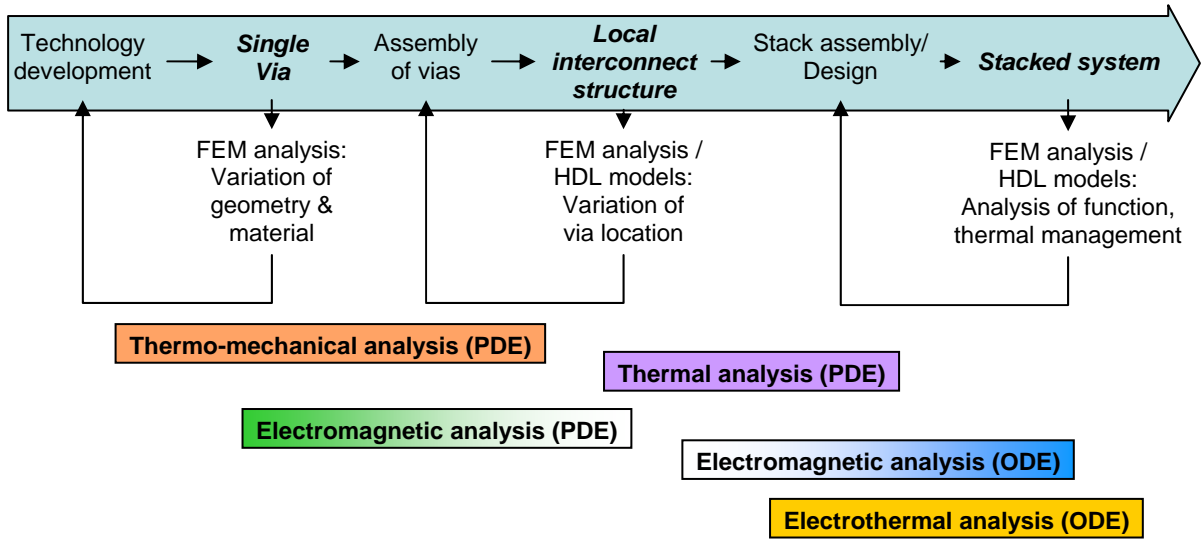

*Figure 2: Modular, hierarchical modeling approach for multi-physics problems in 3D interconnects*

*2. Methods for computer-aided model generation*
The process of model generation for system level simulation is supported by various methods: parameter optimization, model order reduction, and approximation [7], [8]. These methods are combined with direct calculation of model parameters from PDE simulations.

*3. Model validation*
Preparation and measurements of test structures and test circuits are performed in close cooperation with technology development. Due to the dense integration it is necessary to develop special measurement techniques, because devices of interest are often inside the stack and not easily accessible for direct measurements.

*4. Integration of circuit or behavioral models into the design flow*
Models for system level simulation are provided as SPICE macro models or as behavioral models, formulated in description languages as VHDL-AMS, Verilog-AMS, or MAST. This is equivalent to a mathematical description with ordinary differential equations (ODE).

### 3. MODELING TASKS

In the following some aspects of the above mentioned methodology are discussed:

### 3.1. Calculation of circuit parameters

For parameter estimation of parasitic circuit elements, field calculations using PDE solvers are an important method. Using the results of electrostatic and electromagnetic calculations, the values of inductors, capacitors, and resistors can be determined. Models which represent different geometrical detail levels for single vias (see Fig. 3) are created and can be combined to build up typical interconnect structures [9]. Models at higher detail levels are more accurate but even more complex and thus they consume even more memory and simulation time. Detail level 1, the simplest one, is characterized by a uniform via and homogeneous regions for connecting the via to the next stack and for the metallization top of the substrate. Level 2 and 3 reproduce the layer composition of vias are shown. The parameters for coupling $R_k$, $L_k$, $C_k$, and $G_k$ are determined as a function of the distance $x$ of the vias. For multi-via structures a multitude of calculations has to be carried out to obtain the data for parameter estimation.

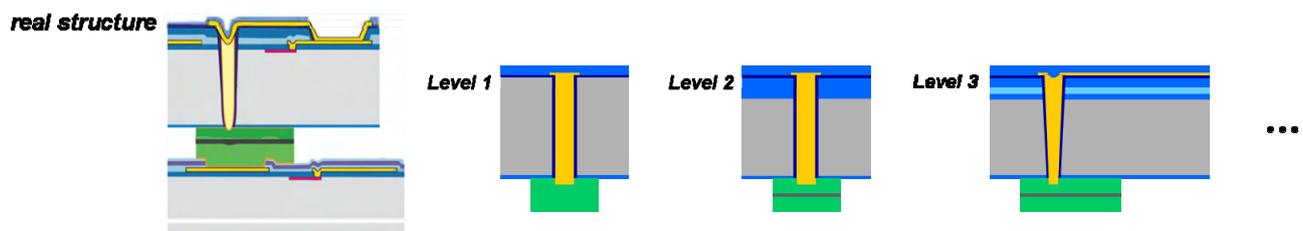

*Figure 3: Models of a single via on different detail levels*

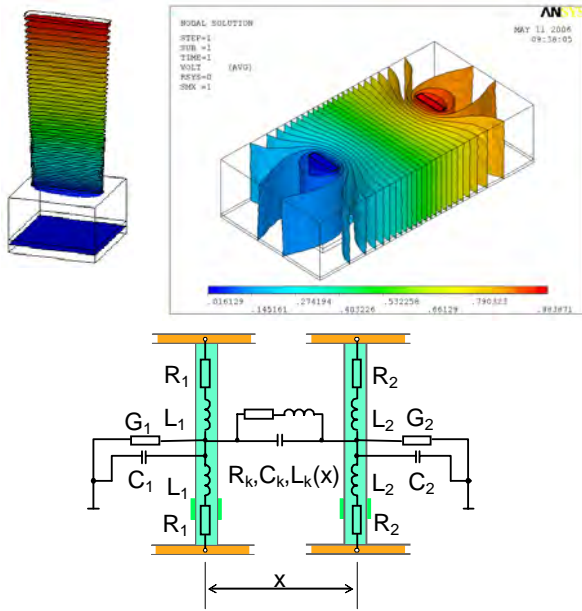

*Figure 4: From left top: finite element model to calculate the resistance of a via; results used to derive the capacitance between two vias; circuit model of two adjacent vias*

### 3.2. Thermal behavior and thermal electrical coupling

Thermal effects play an increasing role in micro systems and integrated circuits [10]. The extension of transistor models towards the consideration of temperature changes includes the introduction of the local device temperature as a variable in all relevant electrical equations. Therefore, the circuit simulator has to fulfill some requirements, especially concerning integration of new (or extended) device models.

Furthermore, it is necessary to calculate the power loss of the devices depending on the electrical properties. This leads to a network representation of the system which allows to consider the coupling of the thermal and electrical behavior. Combined with a thermal network or behavioral model of the stacked structure, a coupled electro-thermal analysis of the entire system is possible. Fig. 5 shows the basic flow of this type of analysis.

To generate the thermal network of the stack, at first the stack structure has to be modeled using the PDE (finite element) simulator ANSYS. The entire model is built up of basic modules which are generated for ANSYS using a common geometrical and material representation (see Fig. 6). In the next step the thermal these two regions more in detail: the vertical shape of the via and the complex structure of the metallization layer is taken into account. Further detail levels are possible but often consume too much memory and simulation time in relation to their benefit. In Fig. 4 results from field calculations and a basic network model for two adjacent network can be derived from the finite element model by order reduction methods directly or by parameter optimization or approximation algorithms using the thermal simulation results of ANSYS.

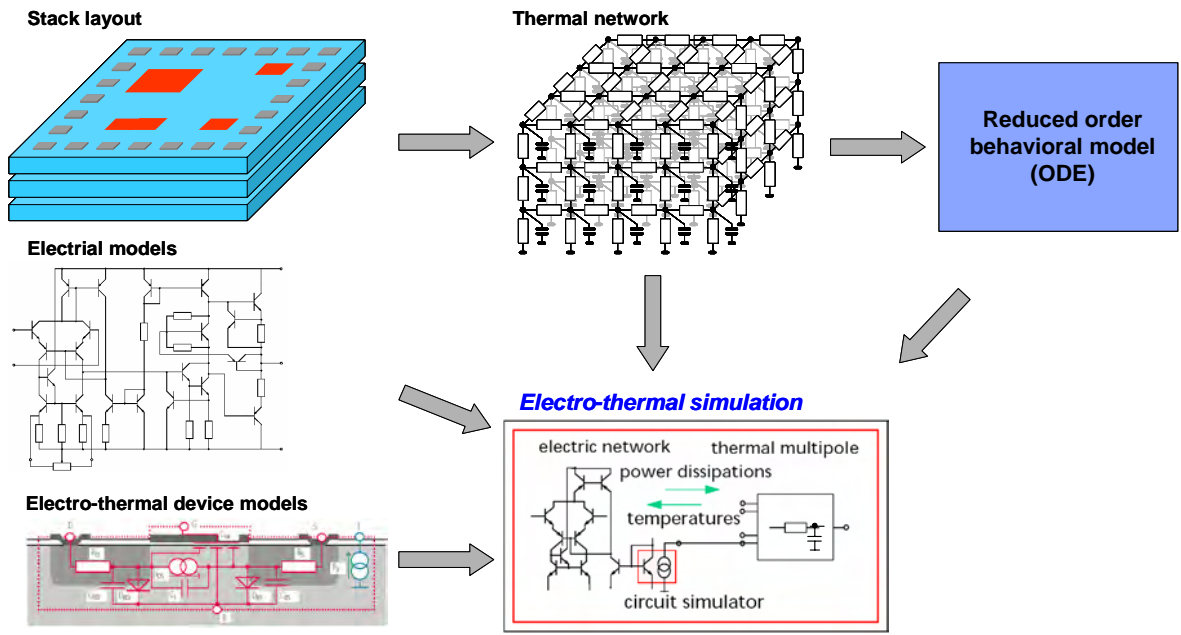

*Figure 5: Basic flow for electro-thermal simulation*

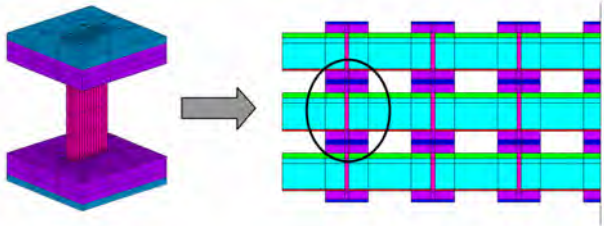

*Figure 6: Thermal simulation: left – basic module of a via, right stack structure built up of basic modules*

Now, the devices of the network representation of the entire stack structure which are temperature-dependent and the devices with power loss have to be replaced by their electro-thermal models.

Finally, the resulting electrical circuit with thermal pins can be connected to the thermal network or to the reduced order behavioral model of the thermal network for shorter simulation times on ODE level.

### 3.3. Electrical behavior at high frequencies

Selected wires and interchip interconnects within a complex stacked system strongly influence the overall system performance. The most important issues, such as EMC and cross talk, are caused by electromagnetic coupling. Timing issues and interconnect delays of distributed blocks can also be investigated [2].

To consider electromagnetic coupling in system design a hierarchical modeling approach is required. First, detailed analyses of structures and physical effects (e.g., skin effect and proximity effect) have to be carried out using PDE solvers. Next, s-parameters, SPICE models, or behavioral models (characterizing the behavior of single and adjacent interconnects) can be derived. These models can be included into the design flow.

### 4. RESULTS OF SELECTED ANALYSES

#### 4.1. Thermal analysis of a three-layer stack

For a stack consisting of three layers a thermal analysis was performed. Due to layout constraints and the request for short wiring between the layers a situation occurs where in one operation mode devices are active at one x-y-location in all three layers. This leads to a local hotspot, shown in Fig. 7 [11].

By rotation of the lower layers each by 90 degree a decrease of maximum temperature by 2 K was achieved,

but a different signal distribution between the layers was required. As an alternative, additional vias for heat transport were investigated [12].

In Fig. 8 the result of a thermal simulation is shown. The maximum temperature decreases to 317 K. Currently the investigations are extended to analyses of additional operation modes and to combination of optimization measures.

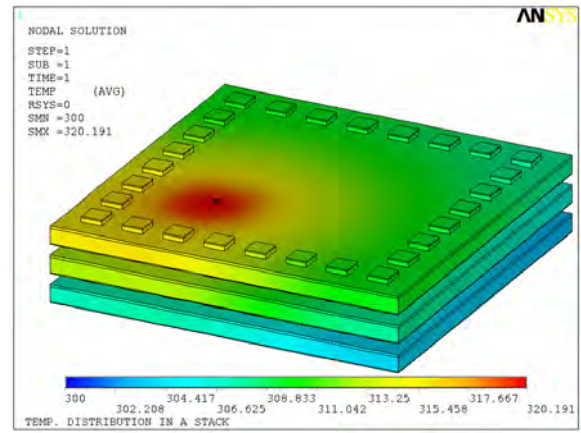

*Figure 7: Stack structure with active devices in the lower left corner, $T_{max}$ = 321 K*

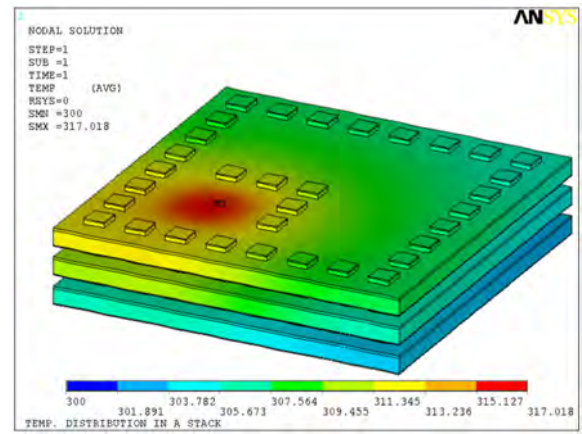

*Figure 8: Stack structure with additional vias, $T_{max}$ = 317 K*

#### 4.2. Analysis of RF behavior of via structures

Especially at high frequencies, the electrical behavior of inter-chip vias is important for the performance of the stacked system [2]. Capacitive and inductive coupling between the vias and to the ground potential as well as the frequency dependence of the via resistance have to be considered. Due to the skin effect the current is mainly flowing at the surface of the conductor. For illustration, the current density of a via with a quadratic cross section at different frequencies is shown in Fig. 9.

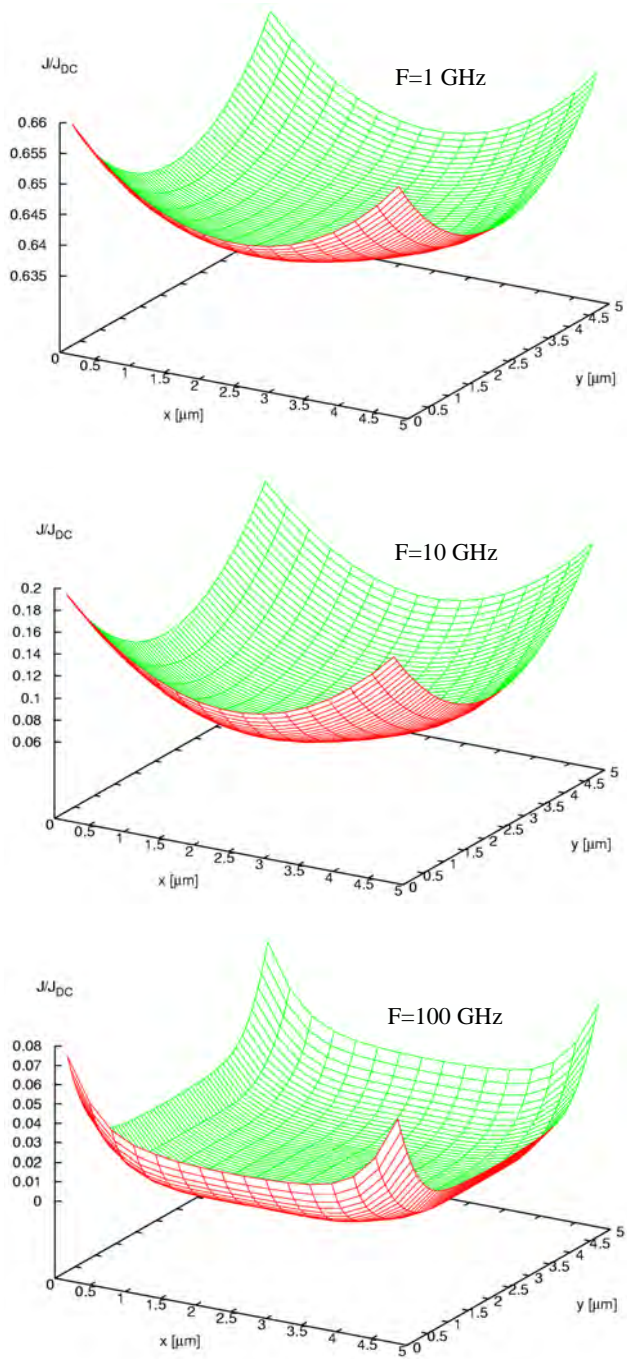

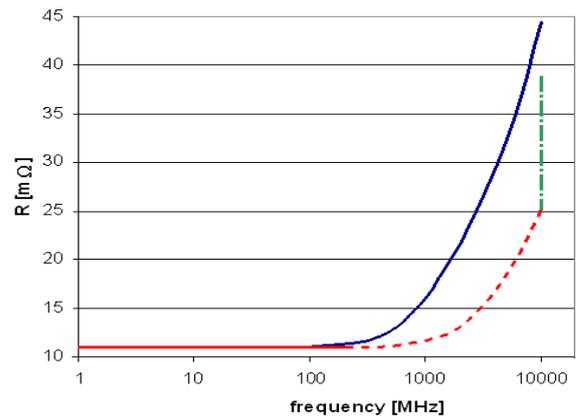

Figure 10: Resistance as a function of frequency:
continuous – single via,
dashed – subdivided vias at fixed distance,
dotted – subdivided vias with decreasing distance

Not only massive circular cross sections of vias are possible but also shapes like shown in Fig. 11. The required area is always the same (7x7 µm²) but the resistance as a function of frequency is partly very different.

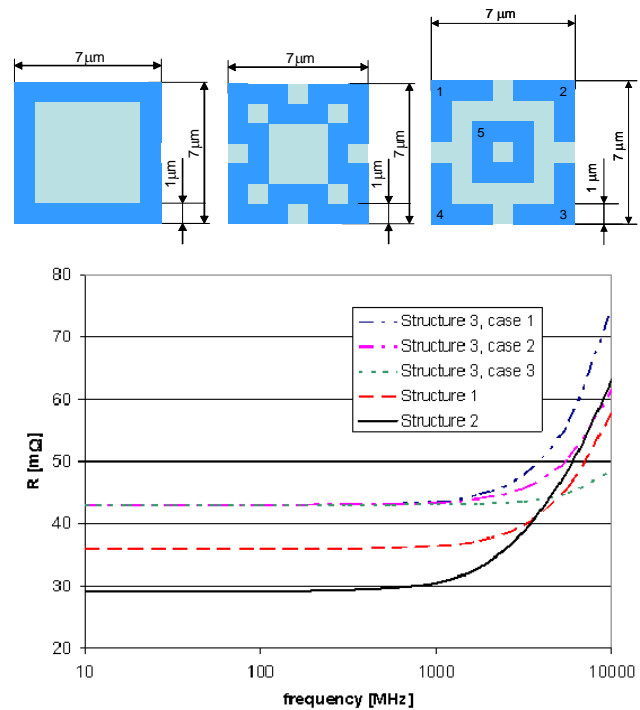

Figure 9: Current density distribution referring to the DC value with the frequency as parameter

In Fig. 10 the increase of the resistance of a via with circular cross section with a diameter of 10 µm is shown (continuous curve). A subdivision into four conductors with the same total area in a distance of 20 µm leads to a minor improvement (dashed curve). A decrease of the distance results in an increase of the resistance (dotted curve) towards the value of the massive conductor.

Figure 11: Top – different cross sections of vias,
below - resulting resistance as a function of the frequency

Structure 1 has a lower effective area than structure 2 and, consequently, a higher DC value of the resistance, but this disadvantage disappears at frequencies higher than approximately 3 GHz (dashed vs. continuous curve). Structure 3 has four outer parallel connected vias

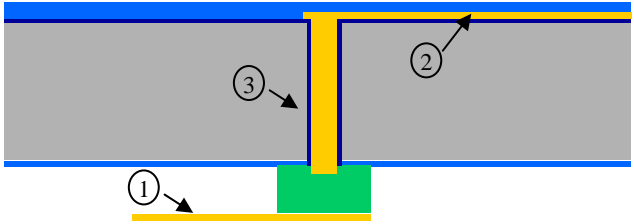

*Figure 12: Transmission lines (1, 2) with via in between (3)*

(1 to 4) and one inner separate via (5), which carries in the first case a current in the same direction like the outer vias. In the second case the inner via is not present and in the last case the current direction in via 5 is opposite to the outer vias. The resulting resistance distributions are shown in Fig. 11. For frequencies higher than about 5 GHz the resistance for case 3 (opposite current direction) becomes lower than the resistance of the structures 1 and 2.

Providing the resistance distributions of the vias as behavioral models they can be used in detailed simulations of the electrical behavior of the system.

In the following example we analyze two transmission lines which are connected with a via (Fig. 12). The lines are represented by simple analytical models and the via is modeled in three ways: at first, using the DC value of the resistance at the entire frequency range; second, using the resistance-frequency-dependency computed with ANSYS (Fig. 10); and last, the via is replaced by a short-circuit.

The resulting transfer behavior of the three cases is shown in Fig. 13. An existing via reduces the cut-off-frequency significantly, but if the resistance is considered as being frequency-dependent, the signal decrease is much more significant at higher frequencies.

## 5. CONCLUSIONS

The goal of all investigations mentioned above is design support for 3D micro systems. Therefore, the adaptation of design flows and the integration of results within these design flows are important tasks, which are also covered by the described methodology. All results are provided using data formats and languages of the dedicated design tools.

Depending on the design task, models on different levels of abstraction can be used. For analog and mixed-signal systems these are usually SPICE net lists or behavioral models programmed in VHDL-AMS or Verilog-AMS. Digital system design is supported by models for crosstalk and signal-dependent delays which are derived from more detailed models and analysis.

Due to the dense integration an important task for the entire stack design is thermal management. This includes the localization of hot spots, adjustment of the stack layout to decrease critical temperatures, as well as reliability issues due to thermally induced mechanical stresses. Furthermore, the influence of thermal effects to the behavior of semiconductor devices and micro structures, e.g. sensor elements, has to be investigated. These tasks are supported either by detailed analysis using PDE solvers or by system level simulation, e.g. for electro-thermal interactions, using appropriate behavioral models.

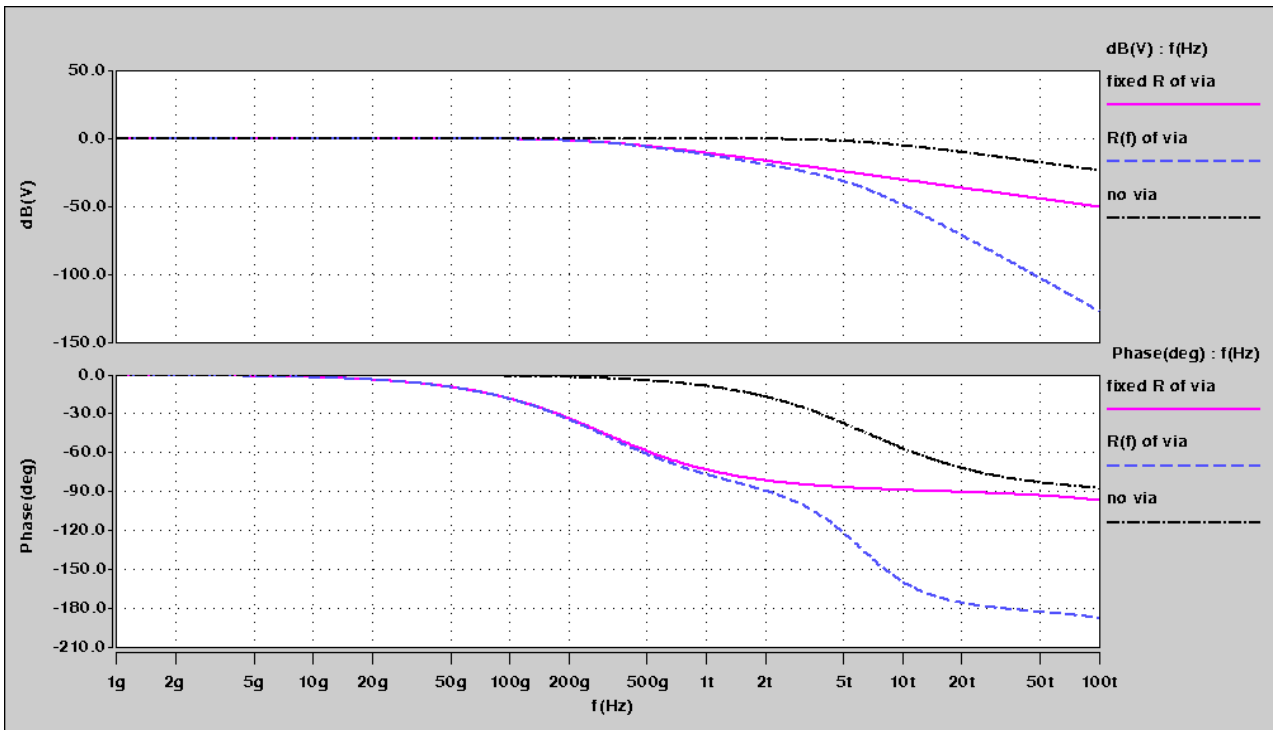

*Figure 13: Transfer behavior for three types of electrical connections: transmission lines with:*
*1 – via with fixed value of its resistance, 2 – via with frequency dependent resistance, 3 – no via between the lines*

## 6. ACKNOWLEDGEMENTS

We would like to express our sincere appreciation to our colleagues at Fraunhofer IZM in Munich and Berlin, especially Josef Weber, Peter Ramm and Eberhard Kaulfersch, and our colleagues at Infineon, Philips and NXP for the highly effective cooperation.

This paper is partly based on the project e-CUBES which is supported by the European Commission under support-no. IST-026461. The authors of this paper are solely responsible for its content.

## 7. REFERENCES

[1] De Man, H.: "Ambient Intelligence: Broad Dreams and Nanoscale Realities", IEEE International Solid-State Circuit Conference ISSCC 2005, San Francisco, February 6-10

[2] Klumpp, A.; Merkel, R.; Weber, J.; Wieland, R.; Elst, G.; Ramm, P.: "Vertical System Integration Technology for High Speed Applications by Using Inter-Chip Vias and Solid-Liquid Interdiffusion Bonding", in "The World of Electronic Packaging and System Integration", ddp goldenbogen, Dresden 2004, pp. 42-47

[3] Benkart, P.; Heittmann, A.; Hübner, H.; Ramacher, U.: "3D Chip Stack Technology using Through-Chip Interconnects", IEEE Design & Test of Computers, 2005, pp. 512-518

[4] Wunderle, B.; Auersperg, J.; Großer, V.; Kaulfersch, E.; Wittler, O.; Michel, B.: Modular parametric finite element modelling for reliability-studies in electronic and MEMS packaging. Proc. DTIP2003, Cannes-Mandelieu, 5-7 Mai 2003

[5] Wunderle, B.; Kaulfersch, E.; Ramm, P.; Michel, B.; Reichl, H.: Thermo-Mechanical Reliability of 3D-integrated Microstructures in Stacked Silicon. Proc. 2006 MRS Fall Meeting, November 27 - December 1, 2006, Boston

[6] Schneider, P. et al.: "Modeling and Simulation". Deliverable report 2006-07-27, Integrated Project e-CUBES, IST-026461, February 2006 - January 2009

[7] Schwarz, P.; Schneider, P.: Model library and tool support for MEMS simulation. Conf. "Microelectronic and MEMS Technology", Edinburgh, Scotland, 2001. SPIE Proc. Series Vol. 4407

[8] Reitz, S.; Bastian, J.; Haase, J.; Schneider, P.; Schwarz, P.: System level modeling of microsystems using order reduction methods. Proc. DTIP 2002, Cannes, Frankreich, pp. 365-373

[9] Elst, G.; Schneider, P.; Ramm, P.: Modeling and Simulation of Parasitic Effects in Stacked Silicon. Proc. 2006 MRS Fall Meeting, November 27 - December 1, 2006, Boston

[10] Shakouri, A.; Kang, S.-M.; Bar-Cohen, A.; Courtois, B. (Editors): Proc. of IEEE, Special Issue On-Chip Thermal Engineering. August 2006, Vol. 94, No. 8

[11] Ababei, C.; Feng, Y.; Golpen, B.; Mogal, H.; Zhang, T.; Bazargan, K.; Sapatnekar, S.: "Placement and Routing in 3D Integrated Circuits", IEEE Design & Test of Computers, 2005, pp. 520-530

[12] Kyu Lim, S.: "Physical Design for 3D System in Package", IEEE Design & Test of Computers, 2005, pp. 532-538